\documentclass[
reprint,
superscriptaddress,
showpacs,preprintnumbers,
amsmath,amssymb,
prb,
]{revtex4-1}

\def\journal #1#2#3#4{#1 {\bf #2}, #3 (#4)}

\usepackage{graphicx}
\usepackage{dcolumn}
\usepackage{bm}
\begin{document}

\title{Modelling of the Peltier effect in magnetic multilayers}
\author{Isaac Juarez-Acosta}
\affiliation{SEPI ENCB, Instituto Politecnico Nacional, Mexico D.F. 11340, Mexico}
\author{Miguel A. Olivares-Robles}
\affiliation{SEPI ESIME Culhuacan, Instituto Politecnico Nacional, Mexico D.F. 04430,
Mexico}
\author{Subrojati Bosu}
\affiliation{Institute for Materials Research, Tohoku University, Sendai 980-8577, Japan}
\affiliation{National Institute for Materials Science, Ibaraki 305-0047, Japan}
\author{Yuya Sakuraba}
\affiliation{National Institute for Materials Science, Ibaraki 305-0047, Japan}
\author{Takahide Kubota}
\affiliation{Institute for Materials Research, Tohoku University, Sendai 980-8577, Japan}
\author{Saburo Takahashi}
\affiliation{Institute for Materials Research, Tohoku University, Sendai 980-8577, Japan}
\author{Koki Takanashi}
\affiliation{Institute for Materials Research, Tohoku University, Sendai 980-8577, Japan}
\author{Gerrit E. W. Bauer}
\affiliation{Institute for Materials Research, Tohoku University, Sendai 980-8577, Japan}
\affiliation{WPI-AIMR, Tohoku University, Sendai 980-8577, Japan}
\affiliation{Kavli Institute of Nanoscience, Delft University of Technology, Lorentzweg
1, 2628 CJ Delft, The Netherlands}
\date{\today}

\begin{abstract}
We model the charge, spin, and heat currents in ferromagnetic metal$|$normal
metal$|$normal metal trilayer structures in the two current model, taking
into account bulk and interface thermoelectric properties as well as Joule
heating. Results include the temperature distribution as well as
resistance-current curves that reproduce the observed shifted parabolic
characteristics. Thin tunneling barriers can enhance the apparent Peltier
cooling. The model agrees with experimental results for wide multilayer
pillars, but the giant effects observed for diameters $%
\lesssim 100\,$nm are still under discussion.
\end{abstract}

\maketitle

\section{Introduction}

Thermoelectric effects result from the coupling between energy and particle
transport in conductors. An example is the heat current that is associated
with a charge current and proportional to a material parameters called
Peltier constant. At a thermocouple, i.e. a junction between two conductors
with different Peltier coefficients, the heat current is not conserved,
which implies heating or cooling depending on the current direction.\cite%
{Snyder2002,Riffat2003} The thermopower, on the other hand, is the
thermoelectric voltage that is generated by a temperature difference over a
metal wire that is proportional to the Seebeck coefficient. A thermocouple
generates an isothermal thermoelectric voltage proportional to the
difference between the Seebeck coefficients when the end of the wires are
maintained at a temperature different from the junction. Nanostructured
materials can enhance the efficiency of thermoelectric devices.\cite%
{Minnich2009} Thermoelectric effects in metallic heterostructures including
ferromagnets depend on the spin degree of freedom.\cite{Silsbee1987} The
spin dependence of thermoelectric cooling\cite%
{Hatami2009,Slachter2010,flipse2012} is part of the field that studies the
coupling between spin, heat and electric transport in small structures and
devices, or spin caloritronics.\cite{bauer2012} Heating is an important
issue for spin torque magnetic random access memories (STT-MRAM) device
[MRAM], and spin caloritronic effects can improve their performance.

An enhanced Peltier effect has been reported by Fukushima \emph{et al}.\cite%
{Fukushima2005,gravier2006,Fukushima2010} in metallic multilayers when
structured into nanopillars. The effect was detected by change $\Delta R$ of
the resistance $R_{0}$ as a function of current bias that acted as a
thermometer. The Peltier effect cools or heats the systems by a term linear
to the applied charge current $I_{c}$ and Peltier coefficient $\Pi$, while
the Joule heating induces a temperature and resistance change that scales
like the square of the current bias such that $\Delta R\sim
R_{0}I_{c}^{2}-\Pi I_{c}$. At small currents the linear term dominates and
causes a reduction of the resistance, i.e. an effective cooling, that in
some structures was found to be very large. The Peltier coefficient was
found by measuring the current where heating and cooling compensate each
other and $\Delta R\left( I_{c}^{\left( 0\right) }\right) =0$ and therefore $%
\Pi=R_{0}I_{c}^{\left( 0\right) }$. The observed $\Pi=480$ mV in pillars
containing Constantan is attractive for cooling nanoelectronic devices.\cite%
{Fukushima2010} The cooling power enhancement was tentatively explained by
Yoshida \textit{et al}.\cite{Yoshida2007,Vu2011} by adiabatic spin-entropy
expansion. However, such an equilibrium cooling mechanism could not explain
that $\Pi$ is material dependent and even changes sign. The diffusion
equation approach by Hatami \textit{et al}.\cite{Hatami2009} did take not
into account either the precise sample configuration nor Joule heating and
could not reproduce the large observed effects. The physical mechanism of
the giant Peltier effect therefore remains unexplained. On the other hand,
the recent experiments by Bosu \textit{et al.}\cite{Bosu2013} confirmed
large Peltier coefficients for pillars including Heusler alloys when
becoming very narrow. The present research has been motivated by the wish to
model the heat and charge currents realistically in the hope to shed light
onto this quandary. We report detailed calculations for the structure and
model parameters matching Bosu \textit{et al.}'s\cite{Bosu2013} experiments
and compare results of semi-analytic calculations with experiments. This
study is limited to thermoelectric effects as described by the two-current
model of thermoelectric transport in which spin current is carried by
particle currents. We do not include explicitly phonon contributions to the
heat current as well as phonon/magnon drag effects on the thermoelectric
coefficients, which may lead to a temperature dependence of the model
parameters. Furthermore, we completely disregarding collective effects that
give rise to e.g. the spin Seebeck and spin Peltier effects.\cite{bauer2012}
There are no indications that these approximations will do more than leading
to some renormalization of the model parameters. While we are still far off
a complete understanding of the experiments, we find evidence that very thin (Ohmic) tunnel
junctions can enhance the Peltier effect.

This paper is organized as follows. In Section II, we review the standard
Valet-Fert model for spin transport\cite{ValetFert1993} in our nanopillars,
with explicit to inclusion of interfaces. In Section III, we extend the
model to include heat currents, charge and spin Joule heating, and explain
our method to compute temperature profiles. In Section IV, we present
results for the Peltier effect due different interfacial thermoelectric
parameters and simulations of the Peltier effect are also performed,
illustrating the importance of interface resistances, to finish in section V
with a summary and conclusions.

\section{Spin-dependent diffusion in $\mathrm{F|N|N}_{\text{\textrm{B}}}$
model}

\begin{figure}[b]
\includegraphics[width=1\columnwidth,clip]{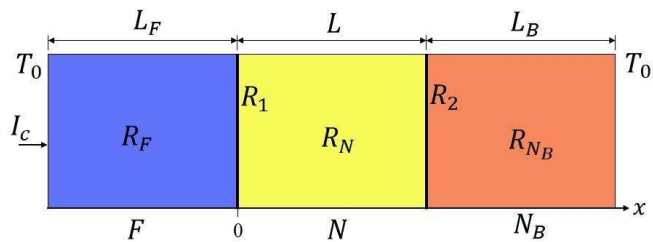}\vspace{0mm}
\caption{(Color online) $\mathrm{F|N|N}_{\text{\textrm{B}}}$ nanopillar
structure biased with a charge current $I_{c}$. We consider a nanopillar
with cross section $A_{c}$. F, N and N$_{\text{\textrm{B}}}$ represent a
ferromagnetic metal, normal metal spacer and normal metal B respectively. L$%
_{\text{\textrm{F}}}$, L and L$_{\text{\textrm{B}}}$ are the thicknesses of
each section. The ends of the nanostructure are connected to thermal
reservoirs kept at a constant temperature $T_{0}$. }
\label{FNNmodel}
\end{figure}

Our model can be applied quite generally to arbitrary multilayered
structures, but we focus here on the charge-current biased trilayer
nanostructures measured by Bosu \emph{et al}.\cite{Bosu2013} that are
composed of a ferromagnetic metal F and two normal metals N and \textrm{N}$_{%
\text{\textrm{B}}}$, respectively, as sketched in Fig. \ref{FNNmodel}. The
thicknesses of F, N and N$_{\text{\textrm{B}}}$ are L$_{\text{\textrm{F}}}$,
L and L$_{\text{\textrm{B}}}$, consecutively, and the device is sandwiched
between two thermal reservoirs at same temperature $T_{0}$. The electric,
spin and heat transport is described by an extended Valet-Fert model,\cite%
{ValetFert1993} including interfaces\cite{Brataas2006} and spin-dependent
thermoelectric effects.\cite{Hatami2009} The parameters are interfaces
resistances $R_{1}$ and $R_{2}$ for interfaces the $\mathrm{F|N}$ and $%
\mathrm{N|N}_{\text{\textrm{B}}}$ respectively,\cite{Son1987,Bass2013} bulk
resistance $R_{i}$ ($i=\mathrm{F,N,N}_{\text{\textrm{B}}}$) for each metal,
as well as the spin polarization $P_{F}$ of the ferromagnetic metal.

We adopt a one-dimensional diffusion model in which the currents flow along
the $x$-direction and the origin is at the $\mathrm{F|N}$ interface. In the
collinear two-channel resistor model, the electrons are in either spin-up or
spin-down states. We divide the structure into various elements such as
resistors, nodes and reservoirs. Discrete resistive elements are interfaces,
tunnel barriers or constrictions that limit the transport. For our purpose,
resistors are separated by nodes in which electrons can be described
semiclassically by distribution functions $f_{i}$. If the interactions
electron-electron or electron-phonon are sufficiently strong, $f_{i}$
approaches the Fermi-Dirac distribution which depends on temperatures $T_{i}$
and chemical potentials $\mu_{i}$. We disregard spin-dependent temperatures
here\cite{Hatami2009} but allow for spin accumulations, \textit{i.e}. local
differences between chemical potentials for both spins.

The spin particle $I_{c}^{(\alpha)}$ and heat $J_{q}^{(\alpha)}$ currents at
a position $x$ in a resistive element are\cite{Hatami2010}
\begin{equation}
I_{c}^{(\alpha)}=A_{c}\int d\epsilon j^{(\alpha)}(\epsilon,x)
\end{equation}
and
\begin{equation}
J_{q}^{(\alpha)}=-\frac{1}{e}\int d\epsilon\epsilon
j^{(\alpha)}(\epsilon,x)-\mu_{0}\int d\epsilon j^{(\alpha)}(\epsilon,x)
\end{equation}
respectively, where $A_{c}$ is the cross sectional area of the nanopillar, $%
\alpha=\uparrow(\downarrow)$ is the electron spin degree of freedom, $%
j^{(\alpha)}$ is the spin, energy ($\epsilon$), and position ($x$) dependent
spectral current density, and $\mu_{0}$ the ground-state chemical potential.
$j^{(\alpha)}=\sigma^{(\alpha)}(\epsilon)f^{(\alpha)}(\epsilon,x)$ is
described by local Fermi-Dirac distributions $f^{(\alpha)}$ at temperature $%
T $ and spin-dependent chemical potentials $\mu^{\alpha}$, times the
energy-dependent conductivity $\sigma^{\alpha}(\epsilon)$.

The spin accumulation is defined as $\mu_{s}=\mu_{\uparrow}-\mu_{\downarrow}$%
, where $\mu_{\uparrow,\downarrow}$ are the spin-up and spin-down chemical
potential of the material, while the charge chemical potential is the
average of the sum of spin-up and spin-down chemical potentials $%
\mu_{c}=(\mu_{\uparrow}+\mu_{\downarrow})/2$.

The transport in each layer is governed by spin and charge diffusion
equations,\cite{ValetFert1993} given by
\begin{equation}
\frac{\partial^{2}}{\partial x^{2}}\mu_{s}=\frac{\mu_{s}}{\lambda^{2}}
\label{spindif}
\end{equation}%
\begin{equation}
\frac{\partial^{2}}{\partial x^{2}}\mu_{c}=-P_{F}\frac{\mu_{s}}{2\lambda^{2}}
\label{chargedif}
\end{equation}
where $\lambda$ is the spin-flip diffusion length, usually much larger in
normal metals than in ferromagnetic metals $\lambda_{N,N_{B}}\gg\lambda_{F}$%
, and
\begin{equation}
P_{F}=\frac{\sigma_{F}^{\uparrow}-\sigma_{F}^{\downarrow}}{\sigma
_{F}^{\uparrow}+\sigma_{F}^{\downarrow}}
\end{equation}
which is the spin polarization of conductivity in terms of the
spin-dependent conductivity for each channel in the ferromagnet. In normal
metals $\mathrm{N}$ and $\mathrm{N}_{\text{\textrm{B}}}$ these polarizations
vanish ($P_{N,N_{B}}=0$). The solution of Eq. (\ref{spindif})
\begin{equation}
\mu_{s}=Y\mathrm{e}^{\frac{x}{\lambda}}+Z\mathrm{e}^{-\frac{x}{\lambda}}
\end{equation}
depends on the integration constants $Y$ and $Z$. For a ferromagnetic metal $%
\mathrm{F}$,\cite{Takahashi20081} we find (see Fig. \ref{FNNmodel})
\begin{equation}
\frac{\mu_{F}^{\left( \alpha\right) }(x)}{e}=-\frac{I_{c}}{\sigma_{F}A_{c}}%
x+\alpha_{1}\pm\frac{\sigma_{F}}{\sigma_{F}^{\alpha}}\beta_{1}\mathrm{e}^{%
\frac{x}{\lambda_{F}}}  \label{mufea}
\end{equation}
where $\alpha_{1}$ is the voltage drop at the interface $\mathrm{F|N}$, $%
\sigma_{F}$ is the electrical conductivity of the ferromagnetic metal, $%
\sigma_{F}^{\alpha}$ the spin-dependent conductivity, $\lambda_{F}$ the
spin-flip diffusion length and $\beta_{1}$ a coefficient to be determined by
boundary conditions at the interface.\cite{Brataas2006}

For the normal metals, the spin-up and spin-down chemical potentials read
\begin{equation}
\frac{\mu_{N}^{\alpha}(x)}{e}=-\frac{I_{c}}{\sigma_{N}A_{c}}x\pm b_{1}%
\mathrm{e}^{-\frac{x}{\lambda_{N}}} \pm b_{2}\mathrm{e}^{\frac{x}{\lambda_{N}%
}}
\end{equation}%
\begin{align}
\frac{\mu_{N_{B}}^{\alpha}(x)}{e} & =-\frac{I_{c}}{\sigma_{N_{B}}A_{c}}%
(x-L)+\alpha_{2}\pm c_{1}\mathrm{e}^{-\frac{x-L}{\lambda_{N_{B}}}}  \notag \\
& \pm c_{2}\mathrm{e}^{\frac{x-L}{\lambda_{N_{B}}}}
\end{align}
for $\mathrm{N}$ and $\mathrm{N}_{\text{\textrm{B}}}$ respectively, where $%
b_{1}$, $b_{2}$, $c_{1}$ and $c_{2}$ complete the number of coefficients
that describe the spin-dependent transport in the present trilayer system.
The spin accumulation in each layer of the $\mathrm{F|N|N}_{\text{\textrm{B}}%
}$ nanowire are $\mu_{s}^{X}(x)$,
while the charge chemical potentials read $\mu_{c}^{X}(x)$,
and the spin-dependent current\cite{Takahashi2008} in a bulk ferromagnetic metal
is (Ohm's Law):
\begin{equation}
I_{X}^{\left( \alpha\right) }(x)=-A_{c}\sigma_{X}^{\left( \alpha\right) }%
\frac{\nabla\mu_{X}^{\left( \alpha\right) }(x)}{e}  \label{spin12}
\end{equation}
where $X=$ $F,N,N_{B}$ and $\sigma_{N}^{\left( \alpha\right) }=\sigma_{N}/2.
$ The spin current $I_{X}^{s}=I_{X}^{\left( \uparrow\right) }-I_{X}^{\left(
\downarrow\right) }$ is the difference between spin-up and spin-down
currents

where parameter such as $R_{\lambda_{X}}=\rho_{X}\lambda_{X}/A_{c}$, which is the resistance over the
spin-flip diffusion length $\lambda_{X}$ in $X$ and $\rho_{X}$ is the
corresponding electrical resistivity, are implicit in the calculations.

\subsection{Interface resistances}

Next we consider spin-dependent transport through the interfaces. We
disregard interface-induced spin-flips,\cite{Bass2013} so at the $\mathrm{F|N%
}$ interface:\cite{Son1987}
\begin{equation}
I_{1}^{\left( \alpha\right) }=\frac{G_{1}^{\left( \alpha\right) }}{e}%
[\mu_{F}^{\left( \alpha\right) }(0)-\mu_{N}^{\left( \alpha\right) }(0)]
\label{eqr1}
\end{equation}
where $G_{1}^{\left( \alpha\right) }$ is the interface conductance with
polarization $P_{1}=\left( G_{1}^{\left( \uparrow\right) }-G_{1}^{\left(
\downarrow\right) }\right) /G_{1}\ $and $G_{1}=G_{1}^{\left( \uparrow
\right) }+G_{1}^{\left( \downarrow\right) }.$ At the interface between the
two normal metals $\mathrm{N|N}_{\text{\textrm{B}}}$
\begin{equation}
I_{2}^{\left( \alpha\right) }=\frac{G_{2}^{\left( \alpha\right) }}{e}%
[\mu_{N}^{\left( \alpha\right) }(L)-\mu_{N_{B}}^{\left( \alpha\right) }(L)]
\label{eqr2}
\end{equation}
Charge $\left( I_{c}=I_{1,2}=I_{1,2}^{\left( \uparrow\right)
}+I_{1,2}^{\left( \downarrow\right) }\right) $ and spin $\left(
I_{1,2}^{s}=I_{1,2}^{\left( \uparrow\right) }-I_{1,2}^{\left(
\downarrow\right) }\right) $ currents are conserved in the interfaces 1 and 2,
and assuming that $R_{1}=1/G_{1}$ and $R_{2}=1/G_{2}$.

\subsection{Boundary conditions}

The boundary conditions are spin and charge current conservation at the
interfaces.
\begin{equation}
I_{F}^{s}(0)=I_{N}^{s}(0)=I_{1}^{s}  \label{bc1}
\end{equation}
for the $\mathrm{F}\mathrm{|N}$ interface
and
\begin{equation}
I_{N}^{s}(L)=I_{N_{B}}^{s}(L)=I_{2}^{s}  \label{bc2}
\end{equation}
for the $\mathrm{N|N}_{\text{\textrm{B}}}$ interface.
We assume that the spin
accumulation vanishes at the end of $\mathrm{N}_{\text{\textrm{B}}}$%
\begin{equation}
\mu_{N_{B}}^{s}(L+L_{B})=0  \label{bc3}
\end{equation}
which is valid for $L_{N}\gg\lambda_{N}$ or $L_{B}\gg\lambda_{N_{B}}$ and/or
when nanopillar diameter widens at $L_{B}$. We can now determine $\beta_{1}$%
, $b_{1}$, $b_{2}$, $c_{1}$ and $c_{2}$ in terms of the coefficients.

Then, it can be now computed the spin
accumulation, spin current and charge chemical potential.

The total electrical resistance, $R=\mu_{c}/\left( eI_{c}\right) $, of the
device can now be written as
\begin{equation}
R=R_{F|N|N_{B}}=R_{F}(x=-L_{F})-R_{N_{B}}(x=L+L_{B})
\end{equation}%
\begin{align}
R(T_{0}) & =-\frac{2P_{F}}{I_{c}(1-P_{F}^{2})}\beta_{1}\mathrm{e}%
^{L_{F}/\lambda_{F}}+\frac{\alpha_{1}}{I_{c}}-\frac{\rho_{F}L_{F}}{A_{c}}
\notag \\
& -\frac{\alpha_{2}}{I_{c}}+\frac{\rho_{N_{B}}L_{B}}{A_{c}}  \label{resist}
\end{align}
where
\begin{equation}
\alpha_{1}=I_{c}R_{1}-\frac{2\beta_{1}(P_{1}-P_{F})}{(1-P_{F}^{2})}%
+P_{1}(b_{1}+b_{2})  \label{alpha1}
\end{equation}
and
\begin{equation}
\alpha_{2}=-I_{c}R_{2}-\frac{I_{c}\rho_{N}L}{A_{c}}  \label{alpha2}
\end{equation}
are the voltage drop at the two interfaces.
\begin{figure}[ptb]
\includegraphics[width=1\columnwidth,clip]{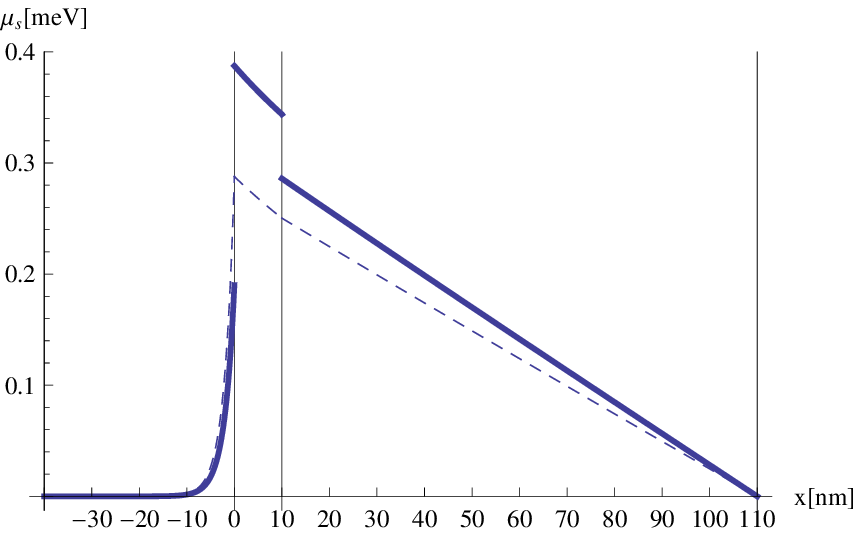}\vspace{0mm}
\caption{(Color online) Spin accumulation in a Co$_{2}$MnSi(CMS)[40nm]$|$%
Au[10nm]$|$Cu[100nm] nanopillar with diameter $D=180$ nm for a current $%
I_{c}=5$ mA and taking interface resistances\protect\cite%
{Sakuraba2010,Henry1996} into account $A_{c}R_{1}=0.915\times10^{-15}\Omega$m%
$^{2}$,\protect\cite{Miura2011} and $A_{c}R_{2}=3.40\times10^{-16}\Omega$m$%
^{2}$ at 300K, the polarization of ferromagnetic metal is $P_{F}=0.71$ and
polarizations of the interfaces\protect\cite{Iwase2009} are $P_{1}=0.77$ and
$P_{2}=0$. The dashed line shows the spin accumulation when interface
resistances $A_{c}R_{1}$ and $A_{c}R_{2}$ are set to zero (metallic
contact). }
\label{spinac}
\end{figure}
\begin{figure}[ptb]
\includegraphics[width=1\columnwidth,clip]{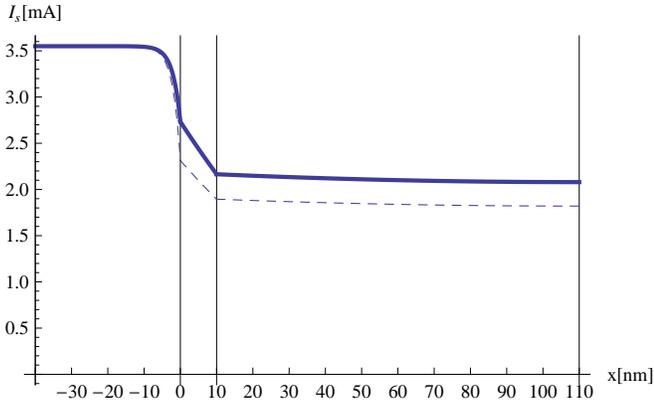}\vspace{0mm}
\caption{(Color online) Spin current in the CMS[40nm]$|$Au[10nm]$|$Cu[100nm]
nanopillar structure for a charge current bias $I_{c}=5$mA and taking
interface resistances into account. The dashed line is the spin current when
interface resistances $A_{c}R_{1}$ and $A_{c}R_{2}$ are set to zero
(metallic contact).}
\label{spincu}
\end{figure}

Numerical results for the transport properties require the parameters of the
samples considered by Bosu \textit{et al.}\cite{Bosu2013} at room
temperature $T_{0}$. The ferromagnetic metal is typically a Heusler alloy Co$%
_{2}$MnSi (CMS),\cite{Nakatani2010} while the normal metal $\mathrm{N}$ is
gold and $\mathrm{N}_{\text{\textrm{B}}}$ is Cu. The resistivities and
spin-flip diffusion lengths are given in Table \ref{materials}.
\begin{table}[ptb]
\begin{ruledtabular}
\begin{tabular}{lcc}
Material&$\lambda$ (nm)&$\rho$ ($\mu\Omega$cm)\\
\hline
Co$_{2}$MnSi&2.1&70.0\\
Au & 60 & 2.27 \\
Cu & 350 & 1.73\\
\end{tabular}
\end{ruledtabular}
\caption{Spin-flip diffusion length and electrical resistivity at 300K used
for the $\mathrm{F|N|N}_{\text{\textrm{B}}}$ nanopillar structure.%
\protect\cite%
{Bosu2013,Nakatani2010,Nishi1987,Ku2006,Yakata2006,Steinhogl2002}}
\label{materials}
\end{table}

Fig. \ref{spinac} illustrates that a charge current $I_{c}$ leads to a spin
accumulation over the spin-flip diffusion length $\lambda_{F}$ in F,
reaching its maximum value at the $\mathrm{F|N}$ interface, where the spin
is injected\cite{Slachter2010,Fert2002,Takahashi2008,Boule2007} and decays
exponentially along the spin-flip diffusion length of the normal metals $%
\lambda_{N,N_{B}}$. The spin current is plotted in Fig. \ref{spincu}. In a
normal metal the spin current is proportional to the gradient of the spin
accumulation, Eq. (\ref{spin12}). It is observed in the model (Fig. \ref%
{spincu}) that the spin current decays rapidly in the central island N. Its
behavior depends strongly on the spin-flip diffusion length of the metal,
for our model we have $\lambda_{Au}<\lambda_{Cu}$. Additionally, it has an
influence from the design length $L,L_{B}$ and the boundary condition
established in Eq. (\ref{bc3}).

\section{Spin-dependent thermoelectricity of $\mathrm{F|N|N}_{\text{\textrm{B%
}}}$ pillars}

In the experiments the electrical resistance change is measured as a
function of applied current, reflecting the balance between the Joule
heating and Peltier cooling. In order to model this effect we need to
compute the temperature profile distribution $T(x)$ over $\mathrm{F|N|N}_{%
\text{\textrm{B}}}$ pillars. Temperature distributions have been previously
calculated, but without taking Joule heating into account in spin-dependent systems.\cite{Hatami2010}
Assuming that we know the temperature dependence of the electrical
resistivity $\rho(T)$ and interface resistances $R_{1,2}\left( T\right) $,
the total temperature dependent resistance reads
\begin{equation}
\Delta R=\frac{1}{L}\int R\left[ T(x)\right] dx-R(T_{0})  \label{chanres}
\end{equation}
where $R(T_{0})$ is given in Eq. (\ref{resist}). For simplicity, we
disregard the heat leaked through the cladding of the nanopillar, which is
valid when the thermal contact is weak or the cladding material has a much
smaller heat conductivity. Significant heat leakage would reduces the
temperature gradients calculated here, leading to an overestimate of the
thermoelectric cooling power. In the following we determine the heat current
and its divergence in the nanopillar taking into account the Kapitza thermal
resistances at interfaces.\cite{Tritt2004} The temperature profile
distribution along the nanopillar structure is calculated using heat
conservation at interfaces, to finally describe the performance of the
nanodevice in the resistance-current ($R$-$I$) characteristics. Except for
the temperature dependence of the resistance that serves as a thermometer,
we disregard the (for elemental metals) weak temperature and voltage
dependences of the thermoelectric parameters.

In the Sommerfeld approximation the linear response relations between
currents and forces in bulk materials read:\cite{Hatami2010}
\begin{equation}
\left(
\begin{array}{c}
J_{c} \\
J_{s} \\
J_{q}%
\end{array}%
\right) =\sigma \left(
\begin{array}{ccc}
1 & P_{F} & ST \\
P_{F} & 1 & P_{F}^{\prime }ST \\
ST & P_{F}^{\prime }ST & \kappa \tau /\sigma%
\end{array}%
\right) \left(
\begin{array}{c}
-\partial _{x}\mu _{c}/e \\
-\partial _{x}\mu _{s}/\left( 2e\right) \\
-\partial _{x}\ln T%
\end{array}%
\right)  \label{Callenequation3}
\end{equation}%
where $S$ is the (charge) Seebeck coefficient, $\sigma $ the electrical
conductivity, $\kappa $ the thermal conductivity, all at the Fermi energy
and $T$ is the temperature (disregarding spin temperatures\cite{Dejene}).
Here, $J_{c}\equiv I_{c}/A_{c}$, etc., are current densities.
\begin{equation}
P_{F}^{^{\prime }}=\frac{\frac{\partial }{\partial E}\left( \sigma
_{F}^{\uparrow }-\sigma _{F}^{\downarrow }\right) _{E_{F}}}{\frac{\partial }{%
\partial E}\left( \sigma _{F}^{\uparrow }+\sigma _{F}^{\downarrow }\right)
_{E_{F}}}  \label{PFprime}
\end{equation}
is the spin polarization of the energy derivative of the conductivity at the
Fermi energy, which is related to the spin polarization of the thermopower
as
\begin{equation}
P_{S}\equiv \frac{S_{\uparrow }-S_{\downarrow }}{S_{\uparrow }+S_{\downarrow
}}=\frac{P_{F}^{^{\prime }}-P_{F}}{1+P_{F}^{^{\prime }}P_{F}}.
\end{equation}%
Joule heating is a source term that causes a divergence in the heat current:%
\cite{callen1948}
\begin{equation}
\frac{\partial }{\partial x}J_{q}=-J_{c}\frac{\partial }{\partial x}\frac{%
\mu _{c}}{e}
\end{equation}%
Including the dissipation due to spin relaxation\cite{Tulapurkar2011,Dejene}
we obtain the matrix expression for the divergence of the current densities%
\begin{equation}
\frac{\partial }{\partial x}\left(
\begin{array}{c}
J_{c} \\
J_{s} \\
J_{q}%
\end{array}%
\right) =\left(
\begin{array}{ccc}
0 & 0 & 0 \\
P_{F} & -\frac{1-P_{F}^{2}}{2\rho \lambda ^{2}} & 0 \\
-J_{c}\frac{\partial }{\partial x} & -J_{s}\frac{\partial }{2\partial x} &
-J_{q}\frac{\partial }{\partial x}%
\end{array}%
\right) \left(
\begin{array}{c}
\mu _{c}/e \\
\frac{\mu _{s}}{2e} \\
T%
\end{array}%
\right) .  \label{Divergence}
\end{equation}

\subsection{Heat currents and temperature profiles in the bulk of the layers}

The divergence of the heat current in the ferromagnet F reads (Eq. (\ref%
{Divergence}))
\begin{equation}
\frac{\partial}{\partial x}J_{q}^{F}=J_{c}^{2}\rho_{F}+\frac{%
(1-P_{F}^{2})\mu_{s}^{2}}{4\rho_{F}\lambda^{2}}+\frac{J_{q}^{2}}{\kappa},
\label{HeatDivergence}
\end{equation}
which equals the derivative of the heat current in Eq. (\ref{Callenequation3}%
)
\begin{align}
\frac{\partial}{\partial x}J_{q}^{F} & =\frac{\partial}{\partial x}\left(
J_{c}S_{F}T-\frac{(P_{F}^{^{\prime}}-P_{F})S_{F}T\mu_{s}}{2\rho_{F}\lambda }%
-\kappa_{F}\frac{\partial}{\partial x}T\right)  \label{HeatDivergence2} \\
& =-\frac{(P_{F}^{^{\prime}}-P_{F})S_{F}T}{\rho_{F}}\frac{\mu_{s}}{%
2\lambda^{2}}-\kappa_{F}\frac{\partial^{2}}{\partial x^{2}}T,
\end{align}
leading to the heat diffusion equation%
\begin{equation}
\frac{\partial^{2}}{\partial x^{2}}T=-\frac{(P_{F}^{^{\prime}}-P_{F})S_{F}T}{%
\rho\kappa}\frac{\mu_{s}}{2\lambda^{2}}-\frac{J_{c}^{2}\rho_{F}^{2}}{%
\rho\kappa}-\frac{(1-P_{F}^{2})\mu_{s}^{2}}{4\rho\kappa\lambda^{2}}- \frac{%
J_{q}^{2}}{\kappa^{2}}.  \label{nablaT}
\end{equation}
Heat transport is carried in parallel by phonons and electrons.\cite%
{Groeneveld1995} We assume here efficient thermalization in and between both
subsystems, meaning that the electron and phonon temperatures are taken to
be identical. The total thermal conductivity then reads $\kappa=\kappa_{e}+%
\kappa_{p}$.
\begin{figure}[t]
\includegraphics[width=1\columnwidth,clip]{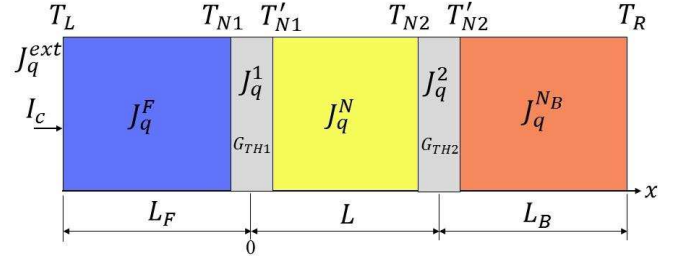}\vspace{0mm}
\caption{(Color online) Definition of temperatures and heat currents in $%
F|N|N_{\text{\textrm{B}}}$ nanopillar structures along the x-direction,
including Kapitza thermal conductances $G_{TH1,2}$. }
\label{tempbo}
\end{figure}

For the ferromagnetic metal F we set $T(x=0)=T_{N1}$ where $T_{N1}$ is
governed by the boundary condition at the $\mathrm{F}\mathrm{|N}$ interface
discussed in the next subsection, while $T(x=-L_{F})=T_{L}$ is fixed by the
reservoir (see Fig. \ref{tempbo}). The solution of the heat diffusion Eq. (%
\ref{nablaT}) disregarding the generilized Thomson effect term $%
-J_{q}^{2}/\kappa^{2}$ then becomes
\begin{align}
T_{F}(x)& =-\frac{2(P_{F}^{^{\prime }}-P_{F})S_{F}T\beta _{1}}{\kappa
_{F}\rho _{F}(1-P_{F}^{2})}[\mathrm{e}^{\frac{x}{\lambda _{F}}}-1]-\frac{%
I_{c}^{2}\rho _{F}x^{2}}{2\kappa _{F}A_{c}^{2}}  \notag \\
& -\frac{\beta _{1}^{2}}{\kappa _{F}\rho _{F}(1-P_{F}^{2})}[\mathrm{e}^{%
\frac{2x}{\lambda _{F}}}-1]+T_{N1}  \notag \\
& +\frac{x}{L_{F}}\left( T_{L}-T_{N1}+\frac{2(P_{F}^{^{\prime
}}-P_{F})S_{F}T\beta _{1}}{\kappa _{F}\rho _{F}(1-P_{F}^{2})}[\mathrm{e}^{%
\frac{x}{\lambda _{F}}}-1]\right.  \notag \\
& \left. +\frac{I_{c}^{2}\rho _{F}x^{2}}{2\kappa _{F}A_{c}^{2}}+\frac{\beta
_{1}^{2}}{\kappa _{F}\rho _{F}(1-P_{F}^{2})}[\mathrm{e}^{\frac{2x}{\lambda
_{F}}}-1]\right)  \label{tempro}
\end{align}%
and
\begin{align}
\frac{\partial }{\partial x}T_{F}& =-\frac{2(P_{F}^{^{\prime
}}-P_{F})S_{F}T\beta _{1}}{\kappa _{F}\rho _{F}(1-P_{F}^{2})}\left( \mathrm{e%
}^{\frac{x}{\lambda _{F}}}{\lambda _{F}}-\frac{[\mathrm{e}^{\frac{L_{F}}{%
\lambda _{F}}}-1]}{L_{F}}\right)  \notag \\
& -\frac{\beta _{1}^{2}}{\kappa _{F}\rho _{F}(1-P_{F}^{2})}\left( \frac{2%
\mathrm{e}^{\frac{2x}{\lambda _{F}}}}{\lambda _{F}}-\frac{[\mathrm{e}^{\frac{%
2L_{F}}{\lambda _{F}}}-1]}{L_{F}}\right)  \notag \\
& -\frac{I^{2}\rho _{F}}{\kappa _{F}A_{c}^{2}}\left( x-\frac{L_{F}}{2}%
\right) +\left( \frac{T_{L}-T_{N1}}{L_{F}}\right).  \label{tritemp}
\end{align}%
Finally, the heat current distribution reads:
\begin{align}
J_{q}\cdot A_{c}& =S_{F}TI_{c}+\frac{2(P_{F}^{^{\prime }}-P_{F})S_{F}T\beta
_{1}}{(1-P_{F}^{2})}\left( \frac{1-\mathrm{e}^{\frac{2L_{F}}{\lambda _{F}}}}{%
R_{F}}\right)  \notag \\
& +\frac{\beta _{1}^{2}}{(1-P_{F}^{2})}\left( \frac{2\mathrm{e}^{\frac{2x}{%
\lambda _{F}}}}{R_{\lambda _{F}}}+\frac{1-\mathrm{e}^{\frac{2L_{F}}{\lambda
_{F}}}}{R_{F}}\right)  \notag \\
& +I_{c}^{2}\left( R_{F}\frac{x}{L_{F}}-\frac{R_{F}}{2}\right) -\frac{\kappa
_{F}A_{c}}{L_{F}}\left( T_{L}-T_{N1}\right) ,  \label{heatferromagnetic}
\end{align}%
where $R_{F}=\rho _{F}L_{F}/A_{c}$ is the electrical and $R_{\lambda
_{F}}=\rho _{F}\lambda _{F}/A_{c}$ the spin resistance.

Repeating this analysis for normal metals, we obtain a heat current in N
\begin{align}
J_{q}^{N}(x)\cdot A_{c}& =S_{N}TI_{c}+I_{c}^{2}\left( R_{N}\frac{x}{L}-\frac{%
R_{N}}{2}\right)  \notag \\
& -b_{1}^{2}\left( \frac{\mathrm{e}^{-\frac{2x}{\lambda _{N}}}}{2R_{\lambda
_{N}}}+\frac{[\mathrm{e}^{-\frac{2L}{\lambda _{N}}}-1]}{4R_{N}}\right)
\notag \\
& +b_{2}^{2}\left( \frac{\mathrm{e}^{\frac{2x}{\lambda _{N}}}}{2R_{\lambda
_{N}}}-\frac{[\mathrm{e}^{\frac{2L}{\lambda _{N}}}-1]}{4R_{N}}\right)  \notag
\\
& -\frac{b_{1}b_{2}}{R_{\lambda _{N}}}\left( \frac{2x-L}{\lambda _{N}}%
\right) -\frac{\kappa _{N}A_{c}}{L}\left( T_{N2}-T_{N1}^{^{\prime }}\right)
\label{heatnormal1}
\end{align}%
and\ $\mathrm{N}_{\text{\textrm{B}}}$%
\begin{align}
J_{q}^{N_{B}}(x)\cdot A_{c}& =S_{N_{B}}TI_{c}+\frac{I_{c}^{2}\rho _{N_{B}}}{%
A_{c}}\left( x-\frac{2L+L_{B}}{2}\right)  \notag \\
& -c_{1}^{2}\left( \mathrm{e}^{-2\frac{x-L}{\lambda _{N_{B}}}}{2R_{\lambda
_{N_{B}}}}+\frac{\mathrm{e}^{-2\frac{L_{B}}{\lambda _{N_{B}}}}-1)}{4R_{N_{B}}%
}\right)  \notag \\
& +c_{2}^{2}\left( \frac{\mathrm{e}^{2\frac{x-L}{\lambda _{N_{B}}}}}{%
2R_{\lambda _{N_{B}}}}-\frac{\mathrm{e}^{2\frac{L_{B}}{\lambda _{N_{B}}}}-1}{%
4R_{N_{B}}}\right)  \notag \\
& -\frac{c_{1}c_{2}}{R_{\lambda _{N_{B}}}}\left( \frac{2x}{\lambda _{N_{B}}}-%
\frac{2L+L_{B}}{\lambda _{N_{B}}}\right) -\frac{\kappa _{N_{B}}A_{c}}{L_{B}}%
\left( T_{R}-T_{N2}^{^{\prime }}\right)  \label{heatnormal2}
\end{align}

\subsection{Interfaces\label{ECI}}

Finally, we knit the solutions for the bulk layers together at the
interfaces by boundary conditions. The contacts to an abruptly widening
nanopillar may be treated as ideal reservoirs (heat and spin sinks) at
constant temperatures $T_{L}=T_{R}=T_{0}$ (see Fig. \ref{tempbo}). By
disregarding interface-induced spin-flips\cite{Bass2013} and, for the
moment, the Joule heating by the interface resistance, we may impose charge,
spin and energy conservation at each interface,\cite%
{gravier2006,Khare2006,Dresselhaus2005} such as $%
J_{q}^{F}(x=0)=J_{q}^{1}=J_{q}^{N}(x=0)$ for F$|$N, where analogous to Eq. (%
\ref{Callenequation3}),\cite{Tritt2004}
\begin{equation}
J_{q}^{1}\cdot A_{c}=G_{\mathrm{TH1}}A_{c}\Delta T-G_{1}S_{1}T_{1}\Delta
\mu_{c}^{(1)}-P_{F}^{^{\prime}}G_{1}S_{1}T_{1}\frac{\Delta\mu_{s}^{(1)}}{2}
\label{Jqinter}
\end{equation}
is the interface heat current, $G_{\mathrm{TH1}}$ the Kapitza thermal
conductance (including the phonon contribution), $A_{c}$ the cross sectional
area of the nanopillar, $\Delta T=T_{N1}-T_{N1}^{\prime}$ the temperature
drop over the interface, $T_{1}=(T_{N1}+T_{N1}^{\prime})/2$ the interface
temperature, $G_{1}$ the electrical interface conductance, $S_{1}$ the
interface thermopower, and $\Delta\mu_{c(s)}^{(1)}$ the charge (spin)
accumulation differences over the interface.

Substituting Eqs. (\ref{heatferromagnetic}) and (\ref{heatnormal1}) for $x=0$
leads to
\begin{align}
T_{N1} & =\left\{ -\left( S_{N}I_{c}+\frac{\kappa_{N}A_{c}}{L}\right) \left(
-\frac{\kappa_{F}A_{c}T_{L}}{H_{2}L_{F}}-\frac{I_{c}^{2}R_{F}}{2H_{2}}%
\right. \right.  \notag \\
& \left. +\frac{\beta_{1}^{2}}{H_{2}(1-P_{F}^{2})}\left( \frac {2}{%
R_{\lambda_{F}}}-\frac{\mathrm{e}^{\frac{2L_{F}}{\lambda_{F}}}-1}{R_{F}}%
\right) \right)  \notag \\
& -\frac{I_{c}^{2}}{2}\left( R_{F}-R_{N}\right) -\frac{\kappa_{F}A_{c}T_{L}}{%
L_{F}}+\frac{\kappa_{N}A_{c}T_{N2}}{L}  \notag \\
& +\frac{\beta_{1}^{2}}{(1-P_{F}^{2})}\left( \frac{2}{R_{\lambda_{F}}}-\frac{%
\mathrm{e}^{\frac{2L_{F}}{\lambda_{F}}}-1}{R_{F}}\right)  \notag \\
& +b_{1}^{2}\left( \frac{1}{2R_{\lambda_{N}}}+\frac{[\mathrm{e}^{-\frac {2L}{%
\lambda_{N}}}-1]}{4R_{N}}\right)  \notag \\
& \left. -b_{2}^{2}\left( \frac{1}{2R_{\lambda_{N}}}-\frac{[\mathrm{e}^{%
\frac{2L}{\lambda_{N}}}-1]}{4R_{N}}\right) +\frac{b_{1}b_{2}}{R_{\lambda_{N}}%
}\left( \frac{-L}{\lambda_{N}}\right) \right\} \diagup  \notag \\
& \left\{ \left( S_{N}I_{c}+\frac{\kappa_{N}A_{c}}{L}\right) \left( \frac{%
S_{F}I_{c}}{H_{2}}+\frac{\kappa_{F}A_{c}}{H_{2}L_{F}}\right. \right.  \notag
\\
& \left. +\frac{2(P_{F}^{^{\prime}}-P_{F})S_{F}\beta_{1}}{H_{2}(1-P_{F}^{2})}%
\left( -\frac{[\mathrm{e}^{\frac{L_{F}}{\lambda_{F}}}-1]}{R_{F}}\right) -%
\frac{H_{1}}{H_{2}}\right)  \notag \\
& -S_{F}I_{c}-\frac{\kappa_{F}A_{c}}{L_{F}}  \notag \\
& \left. -\frac{2(P_{F}^{^{\prime}}-P_{F})S_{F}\beta_{1}}{(1-P_{F}^{2})}%
\left( -\frac{[\mathrm{e}^{\frac{L_{F}}{\lambda_{F}}}-1]}{R_{F}}\right)
\right\}  \label{tn1}
\end{align}
and%
\begin{align}
T_{N1}^{^{\prime}} & =\left( \frac{S_{F}I_{c}}{H_{2}}+\frac{\kappa_{F}A_{c}}{%
H_{2}L_{F}}+\frac{2(P_{F}^{^{\prime}}-P_{F})S_{F}\beta_{1}}{%
H_{2}(1-P_{F}^{2})}\right.  \notag \\
& \left. \left( -\frac{[\mathrm{e}^{\frac{L_{F}}{\lambda_{F}}}-1]}{R_{F}}%
\right) -\frac{H_{1}}{H_{2}}\right) T_{N1}  \notag \\
& -\frac{\kappa_{F}A_{c}T_{L}}{H_{2}L_{F}}-\frac{I_{c}^{2}R_{F_{L}}}{2H_{2}}+%
\frac{\beta_{1}^{2}}{H_{2}(1-P_{F}^{2})}  \notag \\
& \left( \frac{2}{R_{\lambda_{F}}}-\frac{[\mathrm{e}^{\frac{2L_{F}}{%
\lambda_{F}}}-1]}{R_{F}}\right)  \label{tn1p}
\end{align}
where $H_{1(2)}=-G_{1}S_{1}\Delta\mu_{c}^{(1)}/2-P_{F}^{^{\prime}}G_{1}S_{1}%
\Delta\mu_{s}^{(1)}/4\pm G_{\mathrm{TH1}}A_{c}$. We may determine the
temperatures $T_{N2}$ and $T_{N2}^{^{\prime}}$ at interface $\mathrm{N|N}_{%
\text{\textrm{B}}}$ analogously.

Eqs. (\ref{tn1}) and (\ref{tn1p}) include bulk and interfacial Peltier
effects as well as Joule heating in the bulk materials (see Fig. \ref{tempbo}%
) but not yet the interfacial Joule heating.
Here we focus on Joule heating by the $\mathrm{N|N}_{\text{\textrm{B}}}$
interface, which is the dirty one in existing experiments. We can treat
interface heating easily in two limiting cases. In the dirty limit the
interface is a resistor with small but finite thickness $L_{I}$ around the
position $x=d_{I}$ in which the electrons dissipate their energy directly to
the lattice:
\begin{equation}
\frac{\partial }{\partial x}J_{q}^{I}=\left\{
\begin{array}{c}
J_{c}^{2}\frac{R_{I}A}{L_{I}} \\
0%
\end{array}%
\text{ for }%
\begin{array}{c}
-L_{I}/2<x-d_{I}<L_{I}/2 \\
\text{otherwise.}%
\end{array}%
\right.  \label{local}
\end{equation}%
Clean interfaces, point contacts or coherent tunnel junctions, on the other
hand, inject hot electrons (and holes) into the neighboring layers where
they loose their excess energy on the scale of the electron-phonon
thermalization length $\lambda ^{ep}$. In normal metals like Cu it is
surprisingly large even at room temperature, i.e. $\lambda _{Cu}^{ep}=60\,%
\mathrm{nm}$.\cite{Dejene} In the clean limit (assuming that $\lambda
_{A}^{ep}+\lambda _{B}^{ep}$ is smaller than the pillar length)
\begin{equation}
\frac{\partial }{\partial x}J_{q}^{I}=\left\{
\begin{array}{c}
J_{c}^{2}\frac{R_{I}A}{\lambda _{A}^{ep}+\lambda _{B}^{ep}} \\
0%
\end{array}%
\text{ for }%
\begin{array}{c}
-\lambda _{A}^{ep}<x-d_{I}<\lambda _{B}^{ep} \\
\text{otherwise}%
\end{array}%
\right.  \label{delocal}
\end{equation}%
The two limits therefore differ only by the volume in which the heat is
produced. In the extreme case of $\lambda _{A}^{ep}\gg L_{X}$ all interface
Joule heating occurs in the reservoirs, where its effect can be disregarded.
In the following we consider both extremes, i.e. the dissipation occurs
either in the interfacial thickness $L_{I}$ or in the reservoirs $\lambda
_{A}^{ep}+\lambda _{B}^{ep}=\infty $.

We can implement these models into Eqs. (\ref{tn1}) and (\ref{tn1p}) as
follows. In Eqs. (\ref{local}) and (\ref{delocal}), Joule heating is
represented by the power density $J_{c}^{2}R_{I}A/L_{I}$ in the volume $%
V=AL_{I}$. The total power dissipated at the interface is therefore $%
I_{c}^{2}R_{I}.$ This term can be added to Eq. (\ref{tn1}); the first term
of the third line expresses the balance between the Joule heating of the
bulk metals to which the interface contribution may be added. The
interfacial Joule heating thereby reduces the cooling power of the
nanopillar. By contrast, in the ballistic limit and long relaxation lengths
Joule heating is deferred to the heat sinks, and does not contribute at all.
In Eq. (\ref{tn1p}) the interfacial Joule heating is indirectly related by
the already determined term $T_{N1}$ of Eq. (\ref{tn1}). A regular sequence
of the Joule heating is represented by a parabola-like curve, but the
interfacial resistance is a factor of temperature behaviour to result in a
small kink in the temperature distribution at the interface which is
interpreted as bulk heating to be dominant in comparison with the
interfacial one.

\section{Results}

In general, interfacial resistances $R_{1/2}$ may vary from close to zero
for good metallic contacts to that of a very thin (Ohmic) tunnel barrier. A
highly resistive interface can, e.g., be caused by a sample fabrication
process in which the vacuum is broken, leading to organic deposits. We
simulate resistive $\mathrm{F|N}$ or $\mathrm{N|N}_{\text{\textrm{B}}}$
interfaces by modulating $R_{1,2}$ from zero resistance to a large value. A
large resistance of either interface turns out to enhance the cooling effect
as long as the interfacial Joule heating does not dominate, i.e., when the
current bias is not too large.

\subsection{Temperature profiles in a $\mathrm{F|N|N}_{\text{\textrm{B}}}$
pillar}

We are interested in the temperature profile in a pillar with equal
temperatures of the two external reservoirs $T_{L}=T_{R}=T_{0},$ noting that
the model can be easily extended to calculate the thermopower due to a
global temperature difference over the device. We start with $%
T_{N1}=T_{N1}^{\prime }=T_{N2}=T_{N2}^{\prime }=T_{0}$ as initial conditions
(see Fig. \ref{tempbo}), which is substituted into Eqs. (\ref{tn1}) and (\ref%
{tn1p}) to obtain the first iteration. The temperature profiles converge
after several iterations.

\begin{figure}[t]
\includegraphics[width=1\columnwidth,clip]{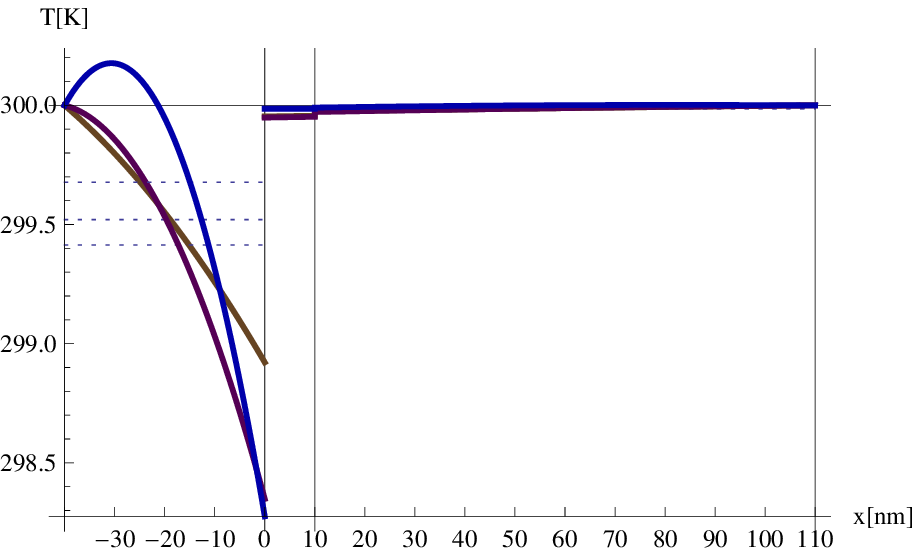}\vspace{0mm} %
\includegraphics[width=1\columnwidth,clip]{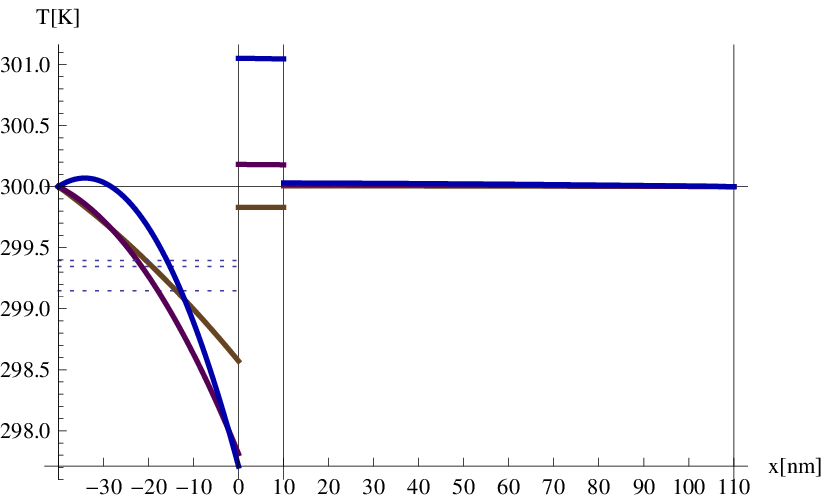}\vspace{0mm}
\caption{(Color online) Temperature distribution in a CMS[40nm]$|$Au[10nm]$|$%
Cu[100nm] nanopillar under current bias for the thermoelectric parameters
from Tables \protect\ref{materials} - \protect\ref{table3}. (Top) Clean $%
\mathrm{F|N}$ interfaces with resistance area of $A_{c}R_{1}=0.915\,\,%
\mathrm{f\Omega m}^{2}$ . The brown, purple and blue lines correspond to
applied current densities of $\ 1,2,3$ times $10^{11}\,\mathrm{A/m}^{2}$,
respectively. The dotted lines for each curve show the average temperature
in the different layers that govern the resistance change of the pillar.
(Bottom) Temperature distribution in the presence of a dirty $\mathrm{N|N}_{%
\text{\textrm{B}}}$ interface with 100 times the electric $A_{c}R_{2}=34\,%
\mathrm{f\Omega m}^{2}$ and heat resistance but otherwise the same
parameters as the clean interface. $T_{0}$ has a constant value of 300K.}
\label{JouleHeating}
\end{figure}

Results for $\mathrm{F|N|N}_{\text{\textrm{B}}}$ nanopillars are shown in
Fig. \ref{JouleHeating} for different current densities, with temperature $%
T_{0}$ in the reservoirs maintained at 300K, using parameters from Tables %
\ref{materials} - \ref{table3} for bulk and interfaces, for the case of all Joule heating occurring in the reservoirs. The top panel of Fig. \ref%
{JouleHeating} is for clean interfaces with $A_{c}R_{1}=0.915\,\mathrm{f}%
\Omega \mathrm{m}^{2}$,\cite{Miura2011} and $A_{c}R_{2}=0.34\,\mathrm{f}%
\Omega \mathrm{m}^{2}$.\cite{Henry1996}. Values of interfacial electrical resistance are well-known parameters, while those of Kapitza heat conductance are not, specially for $F|N$ interfaces, and value of Kapitza heat conductance in latter is assumed not to be such a good heat conductor as compared with the second interface. The
Joule heating is generated mainly by the relatively resistive ferromagnet,
while the cooling takes place at the $\mathrm{F|N}$ interface$\mathrm{,}$
giving rise to a complex temperature and heat current distribution. The
dotted lines for each curve show the average temperature in the different
layers $T_{X_{AVG}}$ that govern the resistance change of the pillar. The
bottom panel of Fig. \ref{JouleHeating} shows the temperature profile in the
presence of a dirty interface $\mathrm{N|N}_{\text{\textrm{B}}}$ with a 100
times larger resistance $A_{c}R_{2}=34\,\mathrm{f}\Omega \mathrm{m}^{2}.$ $%
G_{TH,2}=5.9\cdot 10^{7}\mathrm{W/m}^{2}\mathrm{K}$ is assumed to be reduced
by the same ratio, while other parameters are kept the same. The dissipation
at the dirty interface $\mathrm{N|N}_{\text{\textrm{B}}}$ locally increases
the temperature in the normal metals. A marked discontinuity of the
temperature at $\mathrm{N|N}_{\text{\textrm{B}}}$ interface develops due to
the small thermal conductance $G_{TH,2}$. The temperature on the F-side
drops from approximately 298.3 K for a clean $\mathrm{N|N}_{\text{\textrm{B}}%
}$ interface to 297.7 K for the dirty one (see Fig. \ref{JouleHeating}). The
increased interface resistance forms a barrier for the heat flow from the
heat sinks towards the interface, allowing the region close to the interface
to cool down more efficiently, thereby enhancing the effective Peltier
effect.
\begin{table}[t]
\scalebox{1} {\
\begin{ruledtabular}
\begin{tabular}{lcccc}
Material&$\kappa$ &$S$ &$P_{F}=P_{F}^{'}$& $\partial R_{X}/\partial T$\cite{Bosu2013,Nishi1987,Steinhogl2002}\\
\hline
Co$_{2}$MnSi&15&-20&0.71& $6.07\times10^{-10}\cdot(L_{F}/A_{c})$\\
Au (N) & 318 & 1.83 &  &$8.14\times10^{-11}\cdot(L/A_{c})$\\
Cu (N$_B$) & 401& 1.94& & $6.84\times10^{-11}\cdot(L_{B}/A_{c})$\\
\end{tabular}
\end{ruledtabular}
}
\caption{Thermoelectric parameters of the bulk metal layers in the $\mathrm{%
F|N|N}_{\text{\textrm{B}}}$ nanopillars at 300\thinspace K: Thermal
conductivities $\protect\kappa $ (W/mK),\protect\cite%
{Gundrum2005,Costescu2003,Lyeo2006} and Seebeck coefficients $S$ ($\mathrm{%
\protect\mu }$V/K). $P_{F}$ is the polarization of the conductivity for the
ferromagnet while $P_{F}^{\prime }$ is the polarization of its energy
derivative. For lack of sufficient data we take $P_{F}^{\prime }-P_{F}=0$,
thereby disregarding much of the spin-dependence of the heat diffusion
equations. $R_{X}\left( L_{X}\right) $ are the resistances in $\Omega $ when
thicknesses of the metal layers are in m. }
\label{table2}
\end{table}

\begin{table}[t]
\scalebox{1} {\
\begin{ruledtabular}
\begin{tabular}{lcccc}
Material&$G_{TH}$ &$S$ &$P$ & $\partial R_{X}/\partial T$\cite{Bosu2013,Nishi1987,Steinhogl2002}\\
\hline
CMS$|$Au& $1.8\times10^{8}$ & -4& 0.77 & $(\partial R_{F}/\partial T+\partial R_{F}/\partial T)/2$\\
Au$|$Cu& $5.9\times10^{9}$& 3.5 & 0 & $(\partial R_{N}/\partial T+\partial R_{N_{B}}/\partial T)/2$\\
\end{tabular}
\end{ruledtabular}
}
\caption{Interfacial thermoelectric parameters of the $\mathrm{F|N|N}_{\text{%
\textrm{B}}}$ nanopillars at 300\thinspace K:\protect\cite{Kelly2007}
Interface Kapitza thermal conductances $G_{TH}$ (W/m$^{2}$K) including the
phonon contribution. $S$ ($\mathrm{\protect\mu }$V/K) is the interfacial
Seebeck coefficient and $P$ the spin polarization of the interface
conductance.}
\label{table3}
\end{table}

\subsection{Peltier cooling, Joule heating, and R-I characteristics}

According to Eq. (\ref{chanres}) the temperature profile $T(x)$ is directly
related to the observable resistance change. We compute a specific
temperature profile for a given current bias as sketched below, which can be
used to obtain the total resistance as a function of current that may be
compared with experimental results. To this end we linearize Eq. (\ref%
{chanres}) as:
\begin{equation}
\Delta R_{X}\approx\frac{\partial R_{X}}{\partial T}\left(
T_{X_{AVG}}-T_{0}\right) .
\end{equation}

The total resistance differential is governed by the temperature dependence
of the layer and interface resistances. Each bulk material layer has a
specific $\partial R_{X}/\partial T$, while the calculations establish
average temperatures $T_{X_{AVG}}$ for the sections $\mathrm{F}$, $\mathrm{N}
$ and $\mathrm{N}_{\text{\textrm{B}}}$ respectively, as shown in Fig. \ref%
{JouleHeating} marked by dotted lines. Highly resistive interfaces may might
affect or even dominate the global resistance change when $R_{1(2)}$ and $%
\partial R_{1(2)}/\partial T$ are large. Our calculations include the
temperatures at interfaces $T_{1(2)}$ as expressed in Eq. (\ref{Jqinter}).
For the temperature dependence of the bulk resistivities we adopt the values
listed in Table \ref{table2}. For resistive interfaces we average $\partial
R_{X}/\partial T$ of the two materials; This is expressed in $\mathrm{N|N}_{%
\text{\textrm{B}}}$ interface as represented in Table \ref{table3}:
\begin{equation}
\frac{\partial R_{2}}{\partial T}=\frac{1}{2}\left( \frac{\partial R_{N}}{%
\partial T}+\frac{\partial R_{N_{B}}}{\partial T}\right) ,  \label{tempint}
\end{equation}
while we disregard the temperature dependence of the resistance for good
interfaces.

\begin{figure}[b]
\includegraphics[width=1\columnwidth,clip]{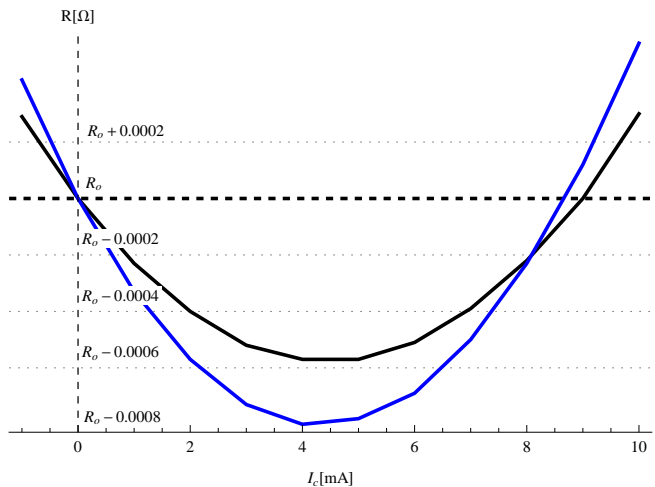}\vspace{0mm}
\caption{(Color online) Resistance-current relation of $CMS$[40\thinspace nm]%
$|Au$[10\thinspace nm]$|Cu$[100\thinspace nm] pillars including interface
resistances as listed in Table \protect\ref{table2}. The effective Peltier
coefficient is $\Pi =R_{0}I_{p}=1.25376\,\Omega \cdot 8.95\,\mathrm{mA}=11.2$%
\thinspace $\mathrm{mV}$ (black line, top). The Peltier coefficient is
increased to $\Pi =R_{0}I_{p}=2.75645\,\Omega \cdot 8.7$\thinspace
\thinspace $\mathrm{mA}=24$\thinspace $\mathrm{mV}$ (blue line, bottom)
when a resistive $\mathrm{N|N}_{B}$ interface of $A_{c}R_{2}=34\,\mathrm{%
f\Omega m}^{2}$ is inserted. For reference, the bulk Peltier coefficient is $%
\Pi _{CMS|Au}=6$\thinspace $\,\mathrm{mV}$. }
\label{ResTun}
\end{figure}

In Fig. \ref{ResTun}, the effect of inserting a highly resistive $\mathrm{N|N%
}_{\text{\textrm{B}}}$ interface on the R-I curves is shown for the scenario
when the interface Joule heating is very non-local, i.e. use Eq. (\ref%
{delocal}). The (effective) Peltier cooling (blue line, bottom) is visibly
enhanced. The change in the total resistance can be understood in terms of
the temperature distribution along the pillar as shown in Fig. \ref%
{JouleHeating}. The increased interfacial resistance $R_{2}$ improves the
effective Peltier coefficient from $\Pi =11.2$\thinspace $\mathrm{mV}$ for a
clean interface to $\Pi =23.9$\thinspace $\mathrm{mV}$ in the case of a
dirty interface. Additionally, a change in the Peltier coefficient from $\Pi
=23.9$\thinspace $\mathrm{mV}$ to $\Pi =24$\thinspace $\mathrm{mV}$ is
reached when Eq. (\ref{tempint}) is implemented into this computation. We should note that while the
effective Peltier coefficient is enhanced by a highly resistive interface
under a constant current bias, it becomes a more efficient system, viz. the nanopillar requires a lower applied
voltage in combination with more cooling effect simultaneously.

\subsection{Trilayer nanopillar model\label{Results}}

We now valuate the thermoelectric performance as a function of structural
and material parameters of the nanopillars. Matching Bosu \textit{et al.}'s%
\cite{Bosu2013} samples, we adopt bulk (Drude) thermopowers of the leads as $%
S_{F}=S_{CMS}=-20\,\mathrm{\mu }$V/K for the ferromagnetic Heusler alloy (Co$%
_{2}$MnSi$_{T_{\mathrm{ann}}=500K}$), $S_{Au}=1.83\,\mathrm{\mu V/K}$ for
the normal metal N and $S_{Cu}=1.94\,\mathrm{\mu V/K}$ in normal metal $%
\mathrm{N}_{\text{\textrm{B}}}$. Our model is scale-invariant with respect
to the pillar diameter, so we cannot explain the enhanced effective Peltier
cooling found in the narrowest pillars by the experiment in terms of an
intrinsic size effect. However, smaller structures can be more susceptible
to the effects of e.g. incomplete removal of resist material used during
nanofabrication. We have disucssed above that such extrinsic effects do
affect the thermoelectric properties and can be treated in our model. The
interfacial thermopower $S_{CMS\mathrm{|}Au}$ and its spin polarization $%
P_{S}$ are basically unknown parameters that may contribute importantly to
the cooling effect in nanostructures, as reflected in the enhancement of the
global effective Peltier coefficient $\Pi =11.2$\thinspace mV for $S_{1}=-4\,%
\mathrm{\mu }$\textrm{V/K} to 23.2\thinspace mV for $S_{1}=S_{CMS|Au}=-30\,%
\mathrm{\mu }$\textrm{V/K}; this case is especially relevant in the presence
of a resistive $\mathrm{N|N}_{\text{\textrm{B}}}$ interface.

\begin{figure}[]
\includegraphics[width=1\columnwidth,clip]{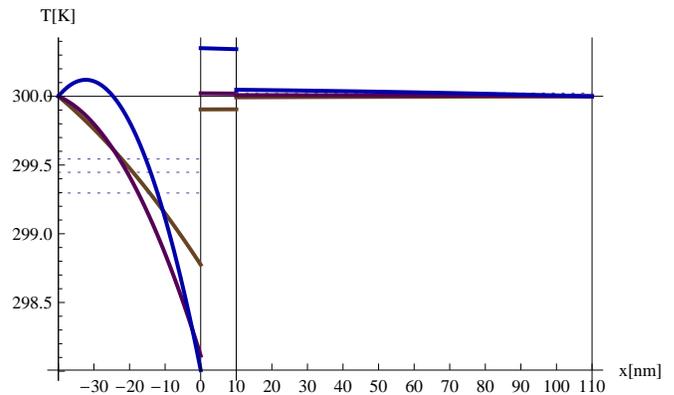}\vspace{0mm} %
\caption{(Color online) Temperature distribution in the presence of a dirty $\mathrm{N|N}_{%
\text{\textrm{B}}}$ interface with 10 times the electric $A_{c}R_{2}=3.4\,%
\mathrm{f\Omega m}^{2}$ and heat resistance but otherwise the same
parameters as the clean interface. $T_{0}$ has a constant value of 300K, setting the nanopillar with the same parameters as Fig. \ref{JouleHeating}, but for the case of Joule heating generated at interfaces instead.}
\label{interfacial}
\end{figure}

The effects of an enhanced interface resistance $A_{c}R_{1(2)}$ on the
Peltier cooling can also be tested by varying it from that of a good
intermetallic to a value corresponding to a thin tunnel barrier. The
interface resistance turns out to improvement of $\Pi $ as long as the
additional Joule heating does not dominate, as illustrated in Fig. \ref%
{JouleHeating}. Furthermore, in Fig. \ref{interfacial} it is plotted the temperature profile distribution when Joule heating is generated in the interfaces, setting the nanopillar with the same parameters of Fig. \ref{JouleHeating}, except for the modulus of the electrical tunnel junction nor the Kapitza thermal conductance, in which both have the same ratio of change. It can be compared clearly a decrement in the performance of the cooling device for this case, since the Joule heating produced at the interface counteracts the cooling of Peltier effect. As discussed above, the interface resistance hinders the flow of heat
current from the heat baths towards the cooling interface. For an
interfacial resistance of $A_{c}R_{1}=0.915\,\,\mathrm{f\Omega m}^{2}$ and $%
A_{c}R_{2}=0.34\,\mathrm{f\Omega m}^{2}$, the total Peltier coefficient
reaches a value of $\Pi _{CMS|Au|Cu}=11.2\,$mV, matching parameters from
Tables \ref{table2} and \ref{table3}, where this result from this
theoretical model is close to experimental ones. A linear dependence of the
Peltier coefficient was found when varying the interface resistance area $%
A_{c}R_{1}$ from $0.915$, $9.15$ and $91.5$ ($\,\mathrm{f\Omega m}^{2}$),
resulting in Peltier coefficients $\Pi _{\mathrm{CMS|Au|Cu}}$ of 11.2,
13.49, and 31.61 mV, respectively. By contrast,when the interfaces are clean
and Joule heating is suppressed (assuming $\lambda _{A}^{ep}+\lambda
_{B}^{ep}=\infty )$, the Peltier coefficients increase to 11.28, 14 and 42mV
for the same interface resistances for the best case when Joule heating is all produced in the reservoirs.

Since our calculations take the spin degree of freedom into account the spin
accumulations and spin currents along the nanopillar are byproducts of the
calculations. In contrast to $\left\vert P_{F}\right\vert <1,$ the spin
polarization of the derivative of the conductivity $-\infty <P_{F}^{^{\prime
}}<\infty .$ When $P_{F}<P_{F}^{^{\prime }}$ the spin contribution to the
cooling power is proportional to the spin accumulations as expressed in Eqs.
(\ref{tn1}) and Eq. (\ref{tn1p}). A Peltier coefficient of $\Pi $ of
11.2\thinspace mV with parameters from Tables \ref{table2} and \ref{table3}
is increased by a factor 2 when $P_{F}^{^{\prime }}=-20$. However, if $%
P_{F}^{^{\prime }}>P_{F}$, the spin degree actually generates heating
thereby reducing the cooling power.

We also studied the dependence of the effective cooling on layer thicknesses
$L_{\text{\textrm{F}}},L$ and $L_{\text{\textrm{B}}}$. The Joule heating
dominates for a critical current bias $I_{c}$ that decreases with increasing
$L_{\text{\textrm{F}}}$. When the thickness of F=CMS is reduced from
40\thinspace nm to approximately 5\thinspace nm, $\Pi $ improves slightly
from 11.2 to 12.5 \textrm{mV}. The optimal thickness of the ferromagnetic
film is $L_{F}\sim \lambda _{F}$. The normal metals do not significantly
contribute to the cooling since their Peltier coefficients are relatively
small.

Finally, slight enhancements of the Peltier coefficient could be achieved by including in the analysis an external heat current $J_{q} ext$, which is depicted in the left hand of Fig. \ref{tempbo}, which forms part of an extension of the nanopillar that could lead the head current towards a further  reservoir so that $T_{L}<T_{0}$, to result in a slight enhancement of the Peltier effect. This makes a more sophisticated model, but we leave it for a future study.

\section{Summary and conclusions}

This paper is motivated by the observed enhancement of the cooling power in
magnetic pillars when the cross section was reduced to the nanoscale.\cite%
{Bosu2013} We develop a realistic spin, charge, and heat diffusion model to
investigate the roles of spin-dependent bulk and interface scattering
contributions. We analyzed the (apparent) cooling power and the conditions
to maximize the effective Peltier effect.

We demonstrate that very thin (Ohmic) tunnel junctions can improve the cooling power of
devices as apparent in the shift of $R(I)$ parabolas. On the other hand, the
spin degree of freedom that was thought to be essential in CMS materials
appears to be less important for conservatively chosen parameters. However,
the material dependence of key parameters is basically unknown. The
parameter $P_{F}^{^{\prime }},$i.e. the spin polarization of the spectral
asymmetry of the conductance, turns out to play an important role. This
parameter may become arbitrarily large when $\left. \partial \left( \sigma
_{F}^{\uparrow }+\sigma _{F}^{\downarrow }\right) /\partial E\right\vert
_{E_{F}}=0$ or, for interfaces $\left. \partial \left( G_{F}^{\uparrow
}+G_{F}^{\downarrow }\right) /\partial E\right\vert _{E_{F}}=0,$ which does
not seem to be an exotic condition and we recommend a systematic search for
such materials or material combinations. Our results also indicate that
interfacial parameters such as the interface Seebeck coefficients $S_{1(2)}$
play a very significant role in the thermoelectric characteristics of
multilayers and may not be disregarded when validating their performance and in agreement with Hu \emph{et al}.\cite{Hu} affirming that the value of this coefficient is even larger that the conventional one, that in the present model both contribute simultaneously in the cooling effect.

While the experiments up now have been analyzed in a simplistic model for
the compensation current at which heating and cooling effects cancel, we
established a distributed model of currents and temperatures. The computed
temperature profiles along the nanopillar established that the cooling is
not homogeneous, but heating and cooling coexists in different locations of
the sample. The current-dependent resistance only a very crude thermometer
that is not a reliable measure for a cooling power that could be of
practical use.

We find that it is possible to selectively cool a ferromagnet by a few
degrees simply by a current flow in the right direction. This could be an
important design parameter for STT-MRAMs. The writing of a bit of
information by a switching event of the free layer in a memory elements is
accompanied by significant Joule and Gilbert heating. Applying a small bias
current after the magnetization reversal can assist a quick return to the
ambient temperature.

Our model is scale invariant with respect to the pillar diameter and does
not provide and intrinsic mechanism for the observed size dependence of the
Peltier effect. In principle, extrinsic effects should exist. The large
fluctuations observed in the experimental results indicates significant
disorder in the smallest nanopillars. One source of the problems can be the
need to break the vacuum during sample fabrication. The effect of pollutants
at an interface are then likely to be more serious for smaller pillars. We
found indeed that by modelling interface as a thin tunnel junction enhances
the apparent Peltier coefficients by suppressing the heat currents flowing
into the pillar from the reservoirs. However, the record cooling effects
observed for some of the narrowest pillars appear to be beyond the effects
that can credibly be modelled, and we cannot exclude the possibility that
something more interesting is going on.

Several effects are beyond the present model approach. Size quantization is
not expected to be important in metallic structures at room temperature, but
could play a role in heterogeneous materials disordered on a
nanometer-scale. Spin waves and magnons, i.e. excitations of the magnetic
order parameter, affect thermoelectric properties. The magnon-drag effect
\cite{Magdrag} enhances the Seebeck coefficient . The longitudinal spin Peltier
effect \cite{spinpeltier} discovered for bilayers with magnetic insulators should
also exist in metallic structures: the spin accumulation in the normal metal
generates a heat current that comes on top of the heat currents discussed
here. It is not clear, however, how and why these effects become so strongly
enhanced in the nanopillars addressed experimentally. More experiments on
even smaller and more reproducibly fabricated nanopillars, preferably
fabricated without breaking the vacuum, are necessary in order to provide
hints on what is going on.

We conclude that the Peltier effect in magnetic nanopillars with diameters $%
\gtrsim 100\,\mathrm{nm}$ appears to be well understood, but that the
enhanced values for narrower ones are to date only partly explained. In
order to employ the large observed effects, more experiments are necessary
in order to shed light on the underlying physical mechanisms.

\begin{acknowledgments}
I. J. A. is grateful to O. Tretiakov, T. Chiba and A. Cahaya for fruitful
discussions and all members of the Bauer Laboratory at the IMR, Tohoku
University for their hospitality. This work was supported by The National
Council for Science and Technology (Scholar reference: 338381), Mexico
(Conacyt), Instituto Politecnico Nacional (REG. SIP 20150488),JSPS\ Grants-in-Aid for Scientific Research (KAKENHI) Nos.
25247056, 25220910, and 26103006, FOM (Stichting voor Fundamenteel Onderzoek
der Materie),\ the ICC-IMR, and DFG Priority Programme 1538
\textquotedblleft {Spin-Caloric Transport}\textquotedblright\ (BA 2954/2).
\end{acknowledgments}


\nocite{*}

\begin{thebibliography}{99}
\bibitem{Snyder2002} G.J. Snyder, J-P. Fleurial and T. Caillat,
\journal{J.
Appl. Phys.}{92}{1564}{2002}.

\bibitem{Riffat2003} S.B. Riffat and X. Ma,
\journal{Appl. Ther.
Eng.}{23}{913-935}{2003}.

\bibitem{Minnich2009} A.J. Minnich, M.S. Dresselhaus, Z.F. Ren and G. Chen, %
\journal{Energy Environ. Sci.}{2}{466-479}{2009}.

\bibitem{Silsbee1987} M. Johnson and R. H. Silsbee
\journal{Phys. Rev.
B}{35}{10}{1987}.

\bibitem{Hatami2009} M. Hatami, G.E.W. Bauer, Q. Zhang and P.J. Kelly, %
\journal{Phys. Rev. B}{79}{174426}{2009}.

\bibitem{Slachter2010} A. Slachter, F.L. Bakker, J.P. Adam and B.J. van
Wees, \journal{Nature Phys. Lett.}{10.1038}{1767}{2010}.

\bibitem{flipse2012} J. Flipse, F.L. Bakker, A. Slachter, F.K. Dejene and
B.J. van Wees, \journal{Nature Nanotechnol. Lett.}{10}{1038}{2012}.

\bibitem{bauer2012} G.E.W. Bauer, E. Saitoh and B.J. van Wees, %
\journal{Nature Mat.}{11}{391-399}{2012}.

\bibitem{MRAM} A.D. Kent and D.C. Worledge, Nature Nanotech. 10, 187 (2015).

\bibitem{Fukushima2005} A. Fukushima, H. Kubota, A. Yamamoto, Y. Suzuki, and
S. Yuasa, \journal{IEEE Trans. Magn.}{41}{2571}{2005}.

\bibitem{gravier2006} L. Gravier, A Fukushima, H, Kubota, A. Yamamoto and S.
Yuasa, \journal{J. Phys. D. Appl. Phys.}{39}{5267}{2006}.

\bibitem{Fukushima2010} A. Sugihara, M. Kodzuka, K. Yakushiji, H. Kubota, S.
Yuasa, A. Yamamoto, K. Ando, K. Takanashi, T. Ohkubo, K. Hono and
A.Fukushima, \journal{Appl. Phys. Express}{3}{0652047}{2010}.

\bibitem{Yoshida2007} H. Katayama-Yoshida, T. Fukushima, V. A. Dinh, and K.
Sato, \journal{Jpn. J. Appl. Phys., Part 2}{46}{L777}{2007}.

\bibitem{Vu2011} N.D. Vu, K. Sato and H.K. Yoshida,
\journal{Appl. Phys.
Express}{4}{015203}{2011}.

\bibitem{Bosu2013} S. Bosu, Y. Sakuraba, T. Kubota, I. Juarez-Acosta, T. Sugiyama, K. Saito, M. A. Olivares-Robles, S. Takahashi, G. E. W. Bauer and K. Takanashi, %
\journal{unpublished}{}{}{2015}.

\bibitem{ValetFert1993} T. Valet and A. Fert,
\journal{Phys. Rev.
B}{48}{10}{1993}.

\bibitem{Brataas2006} A. Brataas, G.E.W. Bauer and P.J. Kelly, %
\journal{Phys. Rep.}{427}{157-255}{2006}.

\bibitem{Son1987} P.C. van Son, H. van Kempen and P. Wyder,
\journal{Phys.
Rev. Lett.}{58}{21}{1987}.

\bibitem{Bass2013} J. Bass, CPP-MR. \journal{arXiv}{}{1305.3848}{2013}.

\bibitem{Hatami2010} M. Hatami, G.E.W. Bauer, S. Takahashi and S. Maekawa, %
\journal{Solid State Commun.}{150}{480-484}{2010}.

\bibitem{Takahashi20081} S. Takahashi, S. Maekawa,
\journal{Sci. Technol.
Adv. Mater.}{9}{014105}{2008}.

\bibitem{Takahashi2008} S. Takahashi and S. Maekawa,
\journal{J. Phys. Soc.
Jpn.}{77}{031009}{2008}.

\bibitem{Nakatani2010} T.M Nakatani, T. Furubayashi, S. Kasai, H. Sukegawa,
Y.K. Takahashi, S. Mitani and K. Hono,
\journal{Appl. Phys.
Lett.}{96}{212501}{2010}.

\bibitem{Miura2011} Y. Miura, K. Futatsukawa, S.Nakajima, K.Abe and
M.SHirai, \journal{Phys. Rev. B}{84}{134432}{2011}.

\bibitem{Nishi1987} Y. Nishi, A. Igarashi and K. Mikagi,
\journal{J. Mater.
Sci. Lett.}{6}{87-88}{1987}.

\bibitem{Ku2006} J.-H. Ku, J. Chang, H. Kim and J. Eom,
\journal{Appl. Phys.
Lett.}{88}{172510}{2006}.

\bibitem{Yakata2006} S. Yakata, Y. Ando, T. Miyazaki and S. Mizukami, %
\journal{Jpn. J. Appl. Phys.}{45}{5A}{2006}.

\bibitem{Steinhogl2002} W. Steinhogl, G. Schindler, G. Steinlesber and M.
Engelhardt, \journal{Phys. Rev. B}{66}{075414}{2002}.

\bibitem{Fert2002} A. Fert, J-M. George, H. Jaffres and G. Faini, %
\journal{J. Phys. D: Appl. Phys.}{35}{2443-2447}{2002}.

\bibitem{Boule2007} O. Boulle, V. Cros, J. Grollier, L. G. Pereira,
C.Deranlat, F. Petroff, G. Faini, J. Barnas and A. Fert,
\journal{Nature
Phys.}{3}{492-497}{2007}.

\bibitem{Sakuraba2010} Y. Sakuraba, K. Izumi, T. Iwase, S. Bosu, K. Saito,
K. Takanashi, Y. Miura, K. Futatsukawa, K. Abe and M. Shirai,
\journal{Phys.
Rev. B}{82}{094444}{2010}.

\bibitem{Henry1996} L.L. Henry, Q. Yang, W-C. Chiang, P. Holody, R. Loloee,
W.P. Pratt, Jr. and J. Bass, \journal{Phys. Rev. B}{54}{17}{1996}.

\bibitem{Iwase2009} T. Iwase, Y. Sakuraba, S. Bosu, K. Saito, S. Mitani and
K. Takanashi, \journal{Appl. Phys. Exp.}{2}{063003}{2009}.

\bibitem{Tritt2004} T.M. Tritt. Thermal conductivity: Theory, properties and
applications. \journal{Kluwer Academic/Plenum Publishers}{}{}{2004}.

\bibitem{Dejene} F. K. Dejene; J. Flipse; G. E. W. Bauer; B. J. van Wees.
Nature Physics \textbf{9}, 636 (2013).

\bibitem{callen1948} H.B. Callen, \journal{Phys. Rev.}{73}{11}{1948}.

\bibitem{Tulapurkar2011} A.A. Tulapurkar and Y. Suzuki,
\journal{Phys. Rev.
B}{83}{012401}{2011}.

\bibitem{Groeneveld1995} R. H. M. Groeneveld and R. Sprik,
\journal{Phys.
Rev. B}{51}{17}{1995}.

\bibitem{Gundrum2005} B. C. Gundrum, D. G. Cahill and R. S. Averback, %
\journal{Phys. Rev. B}{72}{245426}{2005}.

\bibitem{Khare2006} R. Khare, P. Keblinski and A. Yethiraj,
\journal{Int. J.
Heat Mass Transfer}{49}{3401-3407}{2006}.

\bibitem{Dresselhaus2005} R. Yang, G. Cheng and M.S. Dresselhaus, %
\journal{Phys. Rev. B}{72}{125418}{2005}.

\bibitem{Kelly2007} M. Hatami, G. E. W. Bauer, Q. Zhang and P. J. Kelly, %
\journal{Phys. Rev. Lett.}{99}{0666603}{2007}.

\bibitem{Costescu2003} R. M. Costescu, M. A. Wall, D. G. Cahill, %
\journal{Phys. Rev. B}{67}{054302}{2003}.

\bibitem{Lyeo2006} H. Lyeo, D. G. Cahill,
\journal{Phys. Rev.
B}{73}{144301}{2006}.

\bibitem{Magdrag} M. V. Costache, G. Bridoux, I. Neumann \& S. O.
Valenzuela, Nature Materials \textbf{11}, 199 (2012)

\bibitem{spinpeltier} J. Flipse, F. K. Dejene, D. Wagenaar, G. E. W. Bauer,
J. Ben Youssef, B. J. van Wees, Phys. Rev. Lett. 113, 027601 (2014).

\bibitem{Hu} S. Hu, H. Itoh, T. Kimura, NPG Asia Mat. 6 , e127 (2014).

\end{thebibliography}
\end{document}